\documentclass[aps,a4paper,superscriptaddress, nofootinbib, showpacs, 12pt]{revtex4}
\usepackage{amsmath}
\allowdisplaybreaks
\usepackage{amssymb}
\usepackage{latexsym}
\usepackage{amsfonts}
\usepackage{graphicx}
\usepackage{ bm, natbib}
\usepackage{dsfont}
\usepackage{graphicx}
\usepackage{slashed}
\usepackage{ulem}
\usepackage{accents}
\usepackage{bigstrut}
\usepackage{hyperref}
\usepackage{xcolor}

\usepackage{epsfig}
\usepackage{color}

\usepackage{multirow} 
\usepackage{mathrsfs}

\usepackage{comment}

\newcommand{\bea}{\begin{eqnarray}}
\newcommand{\eea}{\end{eqnarray}}

\newcommand{\be}{\begin{equation}}
\newcommand{\ee}{\end{equation}}

\begin{document}

\title{Majorana phases in high-scale mixings unification hypotheses}

\author{S.S. AbdusSalam}
\email{abdussalam@sbu.ac.ir}
\affiliation{Department of Physics, Shahid Beheshti University, Tehran, Islamic Republic of Iran} 
\author{M.Z. Abyaneh}
\email{me_zahiri@sbu.ac.ir}
\affiliation{Department of Physics, Shahid Beheshti University, Tehran, Islamic Republic of Iran}
\author{F. Ghelichkhani}
\email{f.ghelichkhani@mail.sbu.ac.ir}
\affiliation{Department of Physics, Shahid Beheshti University, Tehran, Islamic Republic of Iran}
\author{M. Noormandipour}
\email{mrn31@cam.ac.uk}
\affiliation{Department of Physics, Shahid Beheshti University, Tehran, Islamic Republic of Iran}
\affiliation{DAMTP, University of Cambridge, Wilberforce Road, Cambridge, CB3 0WA, UK}

\begin{abstract} 
For addressing the remarkable difference between neutrino and quark mixings, high-scale mixings relations (HSMR) or unification (HSMU) hypotheses were proposed. These phenomenology frameworks have been explored with respect to bounds from neutrino oscillations and relevant cosmological data. However there are caveats with regards to assessing the hypotheses' compatibility with data in a statistically robust and convergent manner because most analysis employ a few sample of points in model parameters' space. A remedy could be achieved by using Bayesian algorithms for the parameters' space exploration. Using this approach, we made global fits of the HSMU and HSMR models to data and find compatible parameter regions, including for Majorana phases. The posterior samples could be used for studying correlations between neutrino observables and prospects for updates of related experiments. 
\end{abstract}


\maketitle
\section{Introduction} 
\label{intro}
One of the most interesting and challenging discoveries of recent era in particle physics is the neutrino mixing. It is quite different from the quark mixing, which in the standard model (SM) is small. This difference can be explained in various ways, for example via the quark-lepton unification~\cite{Pati:1974yy,Georgi:1974sy,Fritzsch:1974nn}, or via the family or flavour symmetries \cite{Altarelli:2010gt,King:2013eh,Holthausen:2013vba, Araki:2013rkf, Ishimori:2014jwa}, which could also appear in grand unification theories (GUT) \cite{Lam:2014kga}.

 One of the most attractive features of the GUT 
theories \cite{Pati:1974yy,Georgi:1974sy,Fritzsch:1974nn} is the quark-lepton unification. Since quarks and leptons live in a joint representation of the GUT symmetry group, their weak interaction properties are correlated. Hence, in principle one should be able to derive the origin of the small and the large
mixing in the quark and the lepton sectors respectively and also get probable relations between them, in these theories.

Moreover,  there are experimental evidence about the quark-lepton unification. One of them is
the so-called quark-lepton complementarity 
relation 
\cite{Raidal:2004iw} 
\begin{equation}
\label{eq1}
\theta_{12} + \theta_{C}  \approx \frac{\pi}{4},
\end{equation} 
which relates the leptonic mixing angle $\theta_{12}$ and the Cabibbo angle $\theta_C$ and is thought to be a footprint of a high scale quark-lepton unification 
\cite{Raidal:2004iw,Minakata:2004xt,Frampton:2004vw,Kang:2005as}. 
Other observations  suggest~\cite{Abe:2011sj,Adamson:2011qu,Abe:2012tg,Ahn:2012nd,An:2012eh} 
\begin{equation}
\label{eq2}
\theta_{13} \approx  \frac{\theta_{C} }{\sqrt{2}}, 
\end{equation} 
which
can also be a hint to some high scale quark-lepton symmetry in an underlying GUT theory \cite{Antusch:2012fb}.

One way to explain the origin of neutrino and quark mixing is the high scale mixing unification (HSMU) hypothesis, which is studied in details in Refs.\cite{Abbas:2014ala,Abbas:2013uqh,Srivastava:2015tza,Srivastava:2016fhg,Haba:2012ar}. This hypothesis proposes that, at a certain high scale the neutrino and quark mixing angles are identical \cite{Mohapatra:2005gs,Mohapatra:2005pw,Agarwalla:2006dj}. Should the HSMU predictions get verified, it gives a strong hint on the quark-lepton unification or some flavour symmetry or both. Also, a recent parameterisation of the neutrino mixing angles is proposed for both the Majorana and Dirac neutrinos,  called the high scale mixings relation (HSMR), which is inspired by the HSMU hypothesis~\cite{Abbas:2015vba,Abbas-Abyaneh2}.  The neutrino Majorana phases were first addressed in~\cite{BilenkyHosek} and a way to theoretically determine
the Majoran phases is discussed in~\cite{GirardiPetcov}.
Quantitatively, the HSMU hypothesis can be written in the following way:
\begin{equation}
\label{eq3}
\theta_{12}=  \theta_{12}^q ,~~\theta_{13}=  \theta_{13}^q ,~~\theta_{23} =  \theta_{23}^q ,
\end{equation}  
where $\theta_{ij}$ (with $i,j=1,2,3$) are leptonic mixing angles and $\theta_{ij}^q$ 
are the quark mixing angles. The HSMU can be considered a special case of the general mixings relations hypothesis~\cite{Abbas:2015vba} (HSMR) 
\begin{equation}
\label{hsmr}
\theta_{12}=  \alpha_{12}^{k_1}\,\theta_{12}^q ,~~\theta_{13}=  \alpha_{13}^{k_2}\,\theta_{13}^q ,~~\theta_{23} =  \alpha_{23}^{k_3}\,\theta_{23}^q ,
\end{equation}  
where for the analysis here we have chosen the real constants $(k_1, k_2, k_3) = (1,1,1)$. There are
results pointing to the advantage of HSMR over the HSMU. However, here we analyse both scenarios
given the new approach for the parameter space explorations. 

Both HSMU and HSMR were analysed within the framework of the SM extended by  the minimal super-symmetric standard model (MSSM). The beginning point of
working with this hypothesis is to  evolve the quark 
mixing angles from the low scale (mass of the $Z$ boson) to  the 
supersymmetry (SUSY) breaking scale (which we take equal to $2 \textrm{ TeV}$) using the renormalisation-group (RG) evolution of the SM.  
After that, one uses the MSSM RG equations for evolution of 
quark mixing angles from the SUSY breaking scale to the scale of the unification of neutrino and quark mixing, i. e., $M_{ \rm Unification} \sim 2 \times 10^{14} \textrm{ GeV}$, where the quark mixing angles after evolution turn out to be  $\theta_{12}^q = 13.02^\circ$, $\theta_{13}^q=0.17^\circ$, $\theta_{23}^q=2.03^\circ$  and $\delta_{\rm cp}^q=68.925^\circ$. Then, the HSMU hypothesis dictates one to put
the quark mixing angles equal to that of 
the neutrinos at the unification scale, which is also the scale of the dimension-$5$ Weinberg operator. In the next step, using the MSSM RG equations, the leptonic mixing parameters
are run down from the unification scale to the SUSY breaking scale.  And finally, the SUSY breaking scale to the low scale, 
one uses the SM RG equations to run the mixing parameters. 

In order to get reasonable values of neutrino masses, the scale of the dimension-$5$ operator can not be higher than $\mathcal{O}(10^{14}) \textrm{ GeV}$~\cite{Weinbergscale}. In~\cite{Abbas:2015vba} it was shown that results do not change qualitatively when HSMR is analysed at scales up to the GUT scale, $ M_{GUT}=10^{16} \textrm{ GeV}$, within the framework of type-I seesaw mechanism. This is because, most of modifications due to RG runnings happen at scales much lower than the typical seesaw. Thus the results obtained via dimension-$5$ operator and those obtained from type-I seesaw mechanism are not very different over most of the parameters' region.
 
Although neutrinos  could be equally Dirac or Majorana particles in nature and many experiments are going on to test this issue
~\cite{Auger:2012ar,Alessandria:2011rc}, for Dirac mass of neutrinos, the Yukawa couplings seem to be unnaturally small. However, there is a subtle way to explain the smallness of neutrino masses, if they are Majorana particles, called sea-saw mechanism~\cite{Minkowski,GellMann:1980vs,Yanagida:1979as,Glashow:1979nm,Mohapatra:1979ia}. In this work we assume that neutrinos are Majorana particles. The fact that the RG evolution of Majorana neutrinos is extensively studied in the literature~\cite{Antusch:2005gp0,Drees2,Antusch:2005gp,Pluciennik, Babu,Antusch0,Mohapatra:2005gs,Mohapatra:2005pw, Agarwalla:2006dj,Casas:2003kh,Abbas:2014ala,Abbas:2013uqh,Srivastava:2015tza,Casas:1999tg}, facilitates our investigation. 

 For HSMU (HSMR) we have a total number of $5$($9$) free parameters that control the top-down evolution of the neutrino mixing parameters namely, the masses of the three light neutrinos $m_i$ and the  phases: the Dirac phase $\delta$ and the two Majorana phases $\phi_i$. For HSMR we have four additional parameters $\alpha_i$. The HSMU hypothesis predicts the Dirac phase to be equal to the CKM phase, but in HSMR they are taken proportional. The Majorana phases are left as arbitrary parameters since they have no quark counterpart. One chooses these  parameters at the unification scale such that all the mixing parameters at the low scale fall within 
the 3$\sigma$ limit of the neutrino measurements.
For this to work, the chosen masses of 
neutrinos must be quasi-degenerate (QD) and normal hierarchical \cite{Abbas:2014ala} and within SUSY framework, $\tan \beta$ must be large.

From a model-building perspective, as explained in~\cite{Abbas:2015vba}, the HSMR can be realised by using flavour symmetries and seesaw mechanism. The HSMR we have considered has $9$ free parameters to  control the top-down evolution of the neutrino mixing parameters including four $\alpha_i$s. Now, since quark mixing angles are constant and each neutrino angle can change over the period of $0-2\pi$ then, for instance, $\alpha_1$ can vary from zero to $\frac{2\pi}{\theta^q_{12}} = \frac{2\pi}{13.02}$.

In the previous works on HSMU  hypothesis it is demonstrated that one of the mass square differences ($\Delta m^2_{21}$) lies
outside the 3$\sigma$ global range~\cite{Mohapatra:2005gs,Agarwalla:2006dj,Abbas:2014ala}. To bring this parameter into the experimental range one has to take into account the low energy SUSY threshold corrections~\cite{Mohapatra:2005gs,Agarwalla:2006dj,Abbas:2014ala}, without Majorana phases, whose importance for  QD neutrinos is discussed in 
Refs.~\cite{Chun1,Chankowski1,Chun2,Chankowski2} and~\cite{Antusch:2005gp0,Drees2,Antusch:2005gp,Pluciennik, Babu,Mohapatra:2005gs,Mohapatra:2005pw,Abbas:2014ala,Abbas:2013uqh,Srivastava:2015tza,Abbas:2015vba}. 
When one includes Majorana phases, there are parameter space regions where no threshold corrections are necessary. We show that even without taking the threshold corrections into account, all predictions are consistent with the observables.  The threshold corrections are dependent on parameters in the
charged-slepton sector, especially the ratio $R=\frac{M_{\tilde e}}{M_{\tilde \mu, \tilde \tau}}$ and
the wino mass parameter which we set to $1.2$ and $400$ GeV respectively. We use a Bayesian parameter space exploration technique~\cite{Skilling,Feroz,multinest}, applied for HSMU and HSMR studies for the first time. Likewise, the inclusion of Majorana phases as free parameters which lead to agreements with all relevant experimental measurements is addressed here for the first time.

Taking the Majorana phases into account for analysing the neutrinos have important consequences. For example, Ref.~\cite{Abbas:2015vba} studied both HSMU and HSMR  for the Majorana neutrinos without Majorana phases in the CP conserving limit. Based on their study,  while HSMU does not have enough parameter space to accommodate the current observations, the HSMR estimate for the neutrino-less double-beta decay based mass scale, $m_{e\,e}$ is also above the bounds  from the Gerda~\cite{Gerda1,Gerda2} and Kamland~\cite{Kamland} experiments.  The values obtained for this observable in Ref.~\cite{Agarwalla:2006dj} are also at odds with experiments. 
The same can be said about the effective $\beta$-decay neutrino mass $m_{e}$. This observable does not depend on whether the neutrinos are Majorana or Dirac. Previous studies have led to sum of the neutrino masses to be above the cosmological limit~\cite{Abbas:2015vba, Abbas-Abyaneh2,Agarwalla:2006dj} for both Dirac and Majorana neutrinos.
Whereas, including the Majorana phases and doing a statistically convergent parameter space exploration, using an interface code developed for this project together with the public codes REAP~\cite{reap} and MultiNest~\cite{Skilling,Feroz,multinest}, we find  while HSMU is ruled out, HSMR offers parameter regions with the above mentioned observables well within the corresponding experimental limits. 

In short, our work is different with previous works in two major aspects. First, the allowed parameter space of the global fit for the neutrino mixing observables is updated and second, our developed package allows us to explore the parameter space throughly using the Bayesian algorithm, instead of taking low statistics for free parameters at high scale.
Furthermore, we have discovered new strong correlations among different experimental observables for the HSMR  with Majorana phases. These correlations do not exist in the literature and are easily testable in the ongoing and upcoming experiments. 
 
This paper is organised in the following way: In section~\ref{majorana}, we explain the outline of the RG running for the mixing parameters of the Majorana neutrinos, including the Majorana phases.  In section ~\ref{scanning}, we describe the basic concepts used in fitting the HSMU and HSMR neutrino parameters to experimental measurements. We present the results of our analyses  in section~\ref{results} and summarise our work in section~\ref{conclusions}. 

\section{Majorana neutrinos and mixing parameters}
\label{majorana}

The RG evolution of the leptonic mixing parameters for Majorana neutrinos is extensively discussed in the literature both for the case where Majorana phases are zero~\cite{Antusch:2005gp0,Drees2,Antusch:2005gp,Pluciennik, Babu,Mohapatra:2005gs,Mohapatra:2005pw,Abbas:2014ala,Abbas:2013uqh,Srivastava:2015tza} or non-zero~\cite{Agarwalla:2006dj}. The most often studied scenario is the one where, below the unification scale, the Majorana mass term for the left handed neutrinos 
is given by the lowest dimensional operator~\cite{Antusch0}
\begin{equation}\label{eq:Kappa:Babu:1993:1}
 \mathscr{L}_{\kappa} 
 =\frac{1}{4} 
 \kappa_{gf} \, \overline{\ell_\mathrm{L}^\mathrm{C}}^g_c\varepsilon^{cd} \phi_d\, 
 \, \ell_{\mathrm{L}b}^{f}\varepsilon^{ba}\phi_a  
  +\text{H.c.} 
  \;,
\end{equation}
in the SM.  In the MSSM, it is given by
\begin{equation}\label{eq:Kappa-MSSM-s}
\mathscr{L}_{\kappa}^{\mathrm{MSSM}} 
\,=\, \mathscr{W}_\mathrm{\kappa} \big|_{\theta\theta}   +\text{h.c.}
= -\tfrac{1}{4} 
  {\kappa}^{}_{gf} \, \mathds{L}^{g}_c\varepsilon^{cd}
 \mathds{h}^{(2)}_d\, 
 \, \mathds{L}_{b}^{f}\varepsilon^{ba} \mathds{h}^{(2)}_a 
 \big|_{\theta\theta}   +\text{h.c.} \;,
\end{equation}
where $\kappa_{gf}$ has mass dimension $-1$, $\ell_\mathrm{L}^C$ is the charge 
conjugate of a lepton doublet and 
\(a,b,c,d \in \{1,2\}\) are $\mathrm{SU}(2)_\mathrm{L}$ indices.  
The double-stroke letters \(\mathds{L}\) and \(\mathds{h}\)
denote the lepton doublets and the up-type Higgs superfield in the MSSM.  
Using this mass operator, the
neutrino masses are  introduced in a rather model independent way since it does not depend on the underlying mass-term generation mechanism. Our calculations do not depend on the new physics and any model which would have generated the dimension five operator. One such model is the see-saw model, whose RG equations are studied in Ref.~\cite{Antusch:2005gp}.
 In fact, previous studies, for Majorana neutrinos with $\phi_i=0$, have shown that including the type-one see-saw mechanism the predictions do not change in any significant way and that the analysis done with effective dimensional-5 operator is quite robust~\cite{Abbas:2015vba}.

Here we do not elaborate on the calculations for the RG running, which can be found in the above mentioned references. In general, we are interested in the RG evolution of the masses, the mixing
angles and the Dirac and the Majorana phases. The mixing angles and physical phases are described
by the PMNS matrix, which is parameterised as 
\begin{eqnarray}\label{eq:StandardParametrizationU}
 U_{PMNS} & = & V \cdot U,
\end{eqnarray}
where 
\begin{equation}
 V=\left(
 \begin{array}{ccc}
 c_{12} c_{13} & s_{12} c_{13} & s_{13} e^{-i\delta}\\
 -c_{23}s_{12}-s_{23}s_{13}c_{12}e^{i\delta} &
 c_{23}c_{12}-s_{23}s_{13}s_{12}e^{i\delta} & s_{23}c_{13}\\
 s_{23}s_{12}-c_{23}s_{13}c_{12}e^{i\delta} &
 -s_{23}c_{12}-c_{23}s_{13}s_{12}e^{i\delta} & c_{23}c_{13}
 \end{array}
 \right),
\end{equation}
and
\begin{center}
$U= \begin{pmatrix}
 e^{-i\varphi_1/2} & 0&0 \\
   0 & e^{-i\varphi_2/2}&0\\
   0&0&1\\ 
 \end{pmatrix}$,
\end{center}
with \(c_{ij}\) and \(s_{ij}\) defined as \(\cos\theta_{ij}\) and \(\sin\theta_{ij}\) $(i,j = 1,2,3)$, 
respectively. The quantity $\delta$ is the Dirac phase and $\varphi_1$, $\varphi_2$ are the Majorana
phases. In this work, we take Majorana and Dirac 
phases to be nonzero. The non-zero phases are expected to have non-trivial impact on the parameter space. The global experimental status of the leptonic mixing parameter is summarised in Table~\ref{tab:observs}.
\begin{table}[h!]
\begin{center}
\begin{tabular}{|c|c|c|}
  \hline
  PMNS parameters &  Central values  & $3\sigma$ Ranges\\
  \hline
    $\theta_{12}$        & $33.82_{-0.76}^{+0.78}$ & $31.61 - 36.27$       \\
  $\theta_{23}$                       & $49.7_{-1.1}^{+0.9}$ & $40.9 - 52.2$  \\
    $\theta_{13}$                         & $8.61_{-0.13}^{+0.12}$ & $8.22 - 8.98$ \\
    $\delta$                       & $217_{-28}^{+40}$ & $135 - 366$    \\
   $\Delta_{m_{21}}~(10^{-5}~{\rm eV}^2)$        & $7.39_{-0.20}^{+0.21}$ & $6.79 - 8.01$  \\
  $\Delta_{m_{3\ell}}~(10^{-5}~{\rm eV}^2)$        & $2.525_{-0.031}^{+0.033}$ & $2.431 - 2.622$    \\
  \hline
     \end{tabular}
\end{center}
\caption{The central values and $3\sigma$ ranges for neutrino parameters from the global fits~\cite{Globalfit} to experimental measurements. 
  }
 \label{tab:observs}
\end{table}    

For making statistically robust parameter explorations of neutrino models, we developed an interface that combine together the public packages REAP~\cite{reap} (Renormalisation Group Evolution of Angles and Phases)  and MultiNest~\cite{Skilling,Feroz,multinest} via Mathematica++, a library for executing Mathematica from C++ codes and vice versa~\footnote{See \url{www.neelex.com/mathematica++.}}.

 One of the questions since the massiveness of neutrinos was confirmed, has been the scale of the neutrino mass. For QD and normal hierarchy neutrinos spectra, one has
\begin{equation}
m_{1} \lesssim m_{2} \lesssim m_{3} \simeq m_{0}
\end{equation}
with
\begin{equation}
m_{0}
\gg
\sqrt{\Delta{m}^{2}_{32}} \approx 5 \times 10^{-2} \, \text{eV}. 
\label{mQD}
\end{equation}
Three alternative and complementary methods exist to measure the neutrino mass scale.  
The first one is the neutrino-less double beta decay which assumes that neutrinos are Majorana particles \cite{Rodejohann,Bilenky}. The double beta decay effective mass $m_{e\,e}$, is
\be
\label{mee_1}
m_{e\,e}  = \left| \sum U_{ei}^2 \, m_i \right| = \left| m_1 c_{12}^{2} \, 
c_{13}^{2} e^{- i \varphi_1}  + m_2 s_{12}^{2} \, c_{13}^{2}\, 
e^{-i\varphi_2} + m_3 s_{13}^{2} \, e^{-i 2 \delta}
\right| ,
\ee
which for quasi-degenerate neutrinos becomes
\bea
\label{mee_2}
m_{e\,e}& \approx & m_0 \left|  c_{12}^{2} \, 
c_{13}^{2} e^{- i \varphi_1}  + s_{12}^{2} \, c_{13}^{2}\, 
e^{-i\varphi_2} +s_{13}^{2} \, e^{-i 2 \delta}
\right| .
\eea
As a result of the smallness of the 
$\sin^{2}\theta_{13}$ coefficient, the contribution of $m_{3}$ is suppressed and one gets~\cite{Bilenky1,Bilenky2} 
\begin{equation}
\label{quasi1}
m_{e\,e} \simeq m_{0} \, \sqrt{1-\sin^{2}2\theta_{12} \frac{(1-\cos (\varphi_1-\varphi_2))}{2}}.
\end{equation}
which was first derived for the inverted hierarchy neutrino mass spectrum in~\cite{Bilenky1} and for the normal hierarchy case in~\cite{Bilenky2}. A stringent limit on $m_{e\,e}$ comes from the  Gerda~\cite{Gerda1} experiment, $ m_{e\,e}\leq 0.26\, eV$ and an updated limit~\cite{Gerda2} to be  $ m_{e\,e}\leq 0.182\, eV$. The more precise bound comes from Kamland~\cite{Kamland} experiment, $ m_{e\,e}\leq 0.168\, eV$.
The second method to measure the neutrino mass, is to use the kinematics of $\beta$-decay to 
determine the effective electron (anti) neutrino mass ($m_e$), which is a  model independent method and leads to 
\begin{equation} 
\label{mbeta}
m_e \equiv \sqrt{\sum |U_{ei}|^2 \, m_i^2 } \, . 
\end{equation}
The upper bound on $m_e$ from several experiments with tritium beta decay is $2$ eV  \cite{Kraus:2004zw,Aseev:2011dq,Mainz,Drexlin:2013lha}. This has been updated recently to $1.1 eV$ by KATRIN experiment~\cite{KATRIN0}, which should enhance this bound to as low as $0.2$ eV at $90\%$ CL within five years~\cite{KATRIN0}. 

The third method for determining neutrino masses is via cosmological and astrophysical observations, which depends on the cosmological model applied to the data.
The sum of the neutrino masses, $\Sigma m_i$, obtained in this way, has a range for upper bound to be $0.72$ eV at $95\%$ CL \cite{Ade:2015xua}.

\section{Parameter Exploration and Constraints}
\label{scanning}
Using the interface to the REAP and MultiNest packages we explore the parameter
  space of the HSMU and HSMR models. The Bayesian method which is central to the
  exploration algorithm is described as follows. Usually the quantity of interest
  is the Bayesian evidence and/or the model parameters' posterior probability
  distributions. For the model hypothesis, $\cal{H}$, which could be either HSMU or
  HSMR, the parameters at the $M_{\rm Unification}$-scale
\be    
\underline m = \{ \phi_1,  \; \phi_2,  \; m_{1},  \; m_{2},  \; m_{3} \}_{HSMU}
\ee
or
\be    
\underline m = \{ \phi_1,  \; \phi_2,  \; m_{1},  \; m_{2},  \; m_{3},\alpha_{12},\alpha_{13},\alpha_{23},\alpha_{cp} \}_{HSMR}
\ee
are chosen to be in the ranges specified on Tables~\ref{tab:params} and~\ref{tab:params_hsmr}.
\begin{table}[h!]
  \begin{center}
    \begin{minipage}[b]{0.48\hsize}\centering      
      \begin{tabular}{|c|l|}
        \hline
        HSMU Parameter & Range\\
        \hline
        $\phi_{i=1,2}^{[0]}$    & 0 - 360      \\
        $m_{i=1,2,3}$ [eV]     & 0 - 1  \\
        \hline
      \end{tabular}
      \caption{The range of values for the HSMU Majorana neutrino parameters at the unification scale.}
      \label{tab:params}
    \end{minipage}
    \hfill
  \begin{minipage}[b]{0.48\hsize}\centering      
 \begin{tabular}{|ccc|lll|}
        \hline 
        HSMR Parameters &  & &Ranges & & \\
        \hline
        $\phi_{i=1,2}^{[0]}$ & && 0 - 360 & &   \\
        $m_{i=1,2,3}$  [eV] & & & 0 - 1 & &  \\
       $\alpha_{12}$ && & 0 - $\frac{2\pi}{\theta^q_{12}}$ && \\
         $\alpha_{13}$ && & 0 - $\frac{2\pi}{\theta^q_{13}}$ && \\
          $\alpha_{23}$  && & 0 - $\frac{2\pi}{\theta^q_{23}}$ && \\
       
          $\alpha_{CP}$  && & 0 - $\frac{2\pi}{\delta^q_{cp}}$ && \\
        \hline
      \end{tabular}
      \caption{The range of values for the HSMR Majorana neutrino parameters at the unification scale.}
      \label{tab:params_hsmr}
    \end{minipage}
  \end{center}
\end{table}

 Further, all points
within the ranges were taken equally probable with a constant/flat prior probability
distributions, $p(\underline m | {\cal{H}})$. 

The posterior distribution of the parameters in light of the experimental data, $\underline d$,    
considered is given by $p(\underline m| \underline d,{\cal{H}})$. This can be obtained
from Bayes theorem 
\begin{equation} \label{eq.bayes}
p(\underline m|\underline d, {\cal{H}}) = \frac{p(\underline d|\underline
  m,{\cal{H}})p(\underline m|{\cal{H}})}{p(\underline d|{\cal{H}})}.
\end{equation}
Here  
$\mathcal{Z} \equiv p(\underline d|{\cal{H}})$ is the support/evidence for the
model which is defined as the probability density of observing the data set
given that the hypothesis is true.
The evidence is calculated as  
\begin{equation} \label{eq.evid}
\mathcal{Z} =
\int{p(\underline d|\underline m,{\cal{H}}) p(\underline m|{\cal{H}})}\ d \underline m
\end{equation}
where the integral is $N$-dimensional, with $N$ the dimension of the
set of parameters, $\underline m$. $p(\underline d|\underline m, {\cal{H}})$ is  
the likelihood, i.e. the probability of obtaining the data set $\underline d$
given the model parameters, $\underline m$.

We use the MultiNest algorithm~\cite{Feroz,multinest} which implements nested
  sampling technique~\cite{Skilling}. The central values, $\underline \mu$, and 
 corresponding uncertainties, $\underline \sigma$, for the observables  
\begin{eqnarray}    
  \underline O =  \{ \theta_{12}, \; \theta_{23}, \; \theta_{13}, \; \delta_{CP}, \;
  \Delta_{m_{21}}, \; \Delta_{m_{3\ell}} \} 
\end{eqnarray} make the data set used for fitting the Majorana neutrino parameters, 
\be \label{dat} \underline d = \{ \mu_i, \sigma_i \} \ee where $i = 1, \ldots, 6$ 
labels the observables shown in Table~\ref{tab:observs}. Assuming these are independent, 
the combined likelihood used is    
\be p(\underline d|\underline m, {\cal{H}}) = \prod_{i=1}^{8} \,
\frac{ \exp\left[- (O_i - \mu_i)^2/2 \sigma_i^2\right]}{\sqrt{2\pi
    \sigma_i^2}}. 
\ee

We have also demanded three parameters $m_e$, $m_{ee}$ and $\sum m_i$ to be compatible with the latest experimental results.
Each parameter space point from MultiNest sampling is passed to REAP~\cite{Antusch:2005gp}. 
This is used for solving the neutrino physics renormalisation group equations (RGE).
The SM gauge couplings, quark mixing angles, masses of quarks and charged leptons are ran from the low energy 
scale (mass of the Z boson) to the supersymmetry (SUSY) breaking scale, and then to unification scale using MSSM RGEs. 
At the neutrino see-saw unification scale, the neutrino mixing angles are set to be equal to those of the quarks while
the neutrino masses and Majorana phases are treated as free parameters. The leptonic mixing parameters are then
run from the unification scale to the low scale. These are required to agree with current experimental values
shown in Table~\ref{tab:observs}.


\section{Results}
\label{results}

\begin{table*}[ht!]\begin{center}\renewcommand{\arraystretch}{1.12}
\begin{tabular}{|c|c|c||c|c|}
\hline
Parameter  & HSMU central value& $\sigma$  & HSMR  central value& 95\% BCR range \\
\hline
$m_1$(eV) & $0.137$ & $0.0002$ &$0.067$& $0.057-0.086$ \\
\hline
$m_2$(eV) & $0.138$ & $0.0002$ &$0.07$& $0.062-0.092$ \\
\hline
$m_3$(eV) & $0.169$ & $0.0002$ &$0.095$& $0.087-0.11$ \\
\hline
$\phi_1^{[0]} $& $214.2$ & $6.35$ &$156.1$& $0-328$ \\
\hline
$\phi_2^{[0]} $ & $38.8$ & $2.24$ &$179.2$& $0-360$ \\
\hline
$\alpha_{12}$ &$1$ &$0$ &$4.6$&$0.3$-$14.3$ \\
\hline
$\alpha_{13}$ &$1$ &$0$ &$276.8$&$14.3-981.6$ \\
\hline
$\alpha_{23}$ &$1$ &$0$ &$42.2$&$16.5-109.6$ \\
\hline
$\alpha_{cp}$ &$1$ &$0$ &$3.68$&$2.4-4.9$ \\
\hline
\end{tabular}
\caption{The range of values for free parameters at high scale.}\label{tab:fitvalues1}
\end{center}
\end{table*}
In Table~\ref{tab:fitvalues1} we show the central values and $95\%$ Bayesian confidence region(BCR) intervals, or $1 \sigma$ uncertainty values, for the HSMU and HSMR parameters at high scale. These results are important since, they can be used to constrain the theory space of HSMR and also for leptonic model building purposes. It is worthwhile to mention that, the $\alpha$ parameters quantify the difference between the symmetry breaking terms in the Yukawa Lagrangian of the quark and neutrino sector at high scale~\cite{Abbas:2015vba}, and must be different from unity, unless an additional symmetry is involved.
The values of the neutrino mixing angles at high scale in terms of the corresponding quark mixing angles can be deduced from relation~\ref{hsmr} to be $3.9<\theta_{12_{Unif.}}^\circ<186.18$, $2.43<\theta_{13_{Unif.}}^\circ<166.87$, $33.49<\theta_{23_{Unif.}}^\circ<222.48$ and $165.4<\delta^\circ_{_{Unif.}}<337.7$.

\begin{table*}[ht!]\begin{center}\renewcommand{\arraystretch}{1.12}
\begin{tabular}{|c|c|c||c|c|}
\hline
Parameter  & HSMU-1 central value& $\sigma$  & HSMR-1 central value & 95\% BCR range \\
\hline
\hline
$\phi_1$  & $214$ & $6.35$     & $232.5$ & $ 0-360$ \\
\hline
$\phi_2$  & $38.8$ & $ 2.24$  & $181.7$ & $18.5-348.9$ \\
\hline
$ \delta $  & $ 327 $ & $1.22$   & $ 227.5$ & $166.5-288.9$ \\
\hline
$\Sigma m_i$ (eV) & $ 0.3595$ & $0.0005$       & $0.188 $ & $0.167-0.234$ \\
\hline
$\theta_{12}^\circ$  & $34.4$ &   $0.722$      & $33.8$ &   $32.3-35.3$ \\
\hline
$\theta_{13}^\circ$   & $4.87 $ & $0.044$        & $8.61$ & $8.36-8.86$  \\
\hline
$\theta_{23}^\circ$   & $57.8 $ &    $0.872$  & $ 47.6$ &   $44.3-50.2$\\
\hline
$\Delta m^2_{32}$ $(10^{-3} {\rm eV}^2)$  & $2.38 $  & $0.032$     & $2.52$  & $2.46-2.58$ \\
\hline
$\Delta m^2_{21}$ $(10^{-5} {\rm eV}^2)$  & $ 7.57 $  &$0.204$    & $7.40 $  &$7.01-7.80$ \\
\hline
$m_{e\,e}$ (eV) &  $0.1047 $ & $ 0.0002 $                          &  $0.042 $ & $ 0.019-0.056$ \\
\hline
$m_{e}$ (eV) &  $0.1165$ & $ 0.0002 $                              &  $0.057 $ & $ 0.049-0.073 $ \\
\hline
\rm Lightest $m_\nu$: $m_1 $(eV) &  $0.1165 $ & $ 0.0002$ &  $0.056$ & $0.048 -0.072$ \\
\hline
$m_2$(eV) &  $0.117$ & $ 0.00018$ &  $0.057$ & $0.049 -0.073$ \\
\hline
$m_3$(eV) &  $0.126$ & $ 0.00018$ &  $0.076$ & $0.07 -0.088$ \\
\hline
$J_{cp}$ &  $-0.00036$ & $ 0.00001$ &  $-0.021$ & $-0.035 - 0.003$ \\
\hline
\end{tabular}
\end{center}
\caption{The HSMU and HSMR parameters and derived observable at low scale.}\label{tab:fitvalues}
\end{table*}



\begin{figure}[ht]
   \includegraphics[width=.30\textwidth]{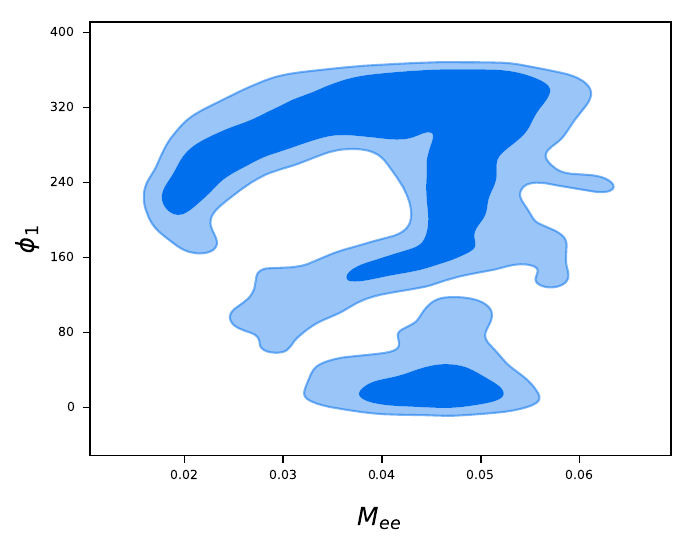}
 \includegraphics[width=.30\textwidth]{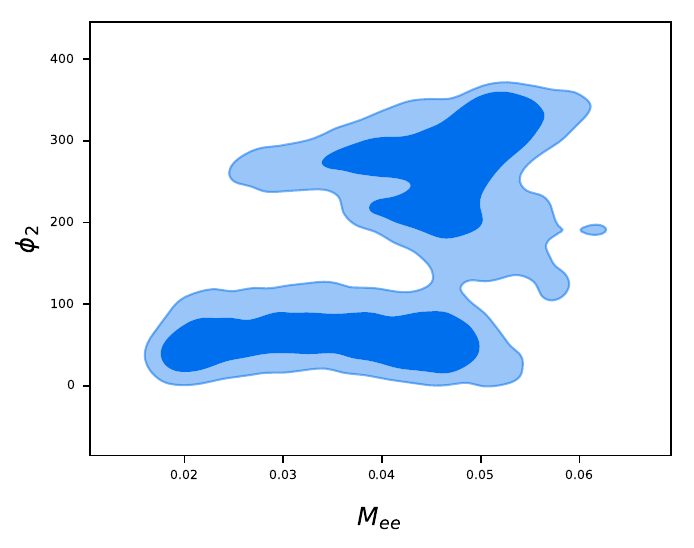}  
  \caption{{Posterior distribution of  Majorana phases with respect to $M_{ee}$ in $\textrm{eV}$ units}.}\label{phimee}
\end{figure}


 In Table~\ref{tab:fitvalues} our predictions for mixing parameters at low scale, including the neutrino masses are given, which  are very precise.
For example, the posterior distributions of  $ m_{e\,e}$ is centered around $ m_{e\,e} \sim 0.042\, eV$ in agreement with Kamland experimental bound. 

Other interesting observables are the absolute neutrino mass, $m_{e}$, and sum of the three neutrino masses, $\Sigma m_i$. The posterior distributions peak around $0.057\, eV$ for $m_{e}$ in accord with the allowed order for the lightest neutrino mass in QD scenarios~(\ref{mQD}) and can be tested by laboratory based neutrino experiments such as KATRIN. Also, $\Sigma m_i < 0.234 eV$ is compatible with cosmological and astrophysical observation~\cite{Ade:2015xua}. The atmospheric neutrino mixing parameter $\theta_{23}$ is non-maximal. 

Among other quantities one could mention the $\Delta m^2_{21}$ and $\Delta m^2_{32}$.
Our results for these quantities are close to the parameter space of the global fit and especially the range of  $\Delta m^2_{21}$ excludes some part of the parameter space of the global fit. Also, our central values nearly coincide with that of the global fit. In the same manner, the range of our results for $\theta_{12}$, $\theta_{13}$ and $\theta_{23}$ excludes a significant part of the parameter space of the $3\-\sigma$ ranges of global fit and the central values for  $\theta_{12}$ and  and $\theta_{23}$ are very close to those of the global fit and that of $\theta_{13}$ coincides with the global fit. 

As in Ref.~\cite{Krastev-cp}, the magnitude of CP violation in 3-flavour neutrino
   oscillations is determined by an invariant which is analogous to the
   Jarlskog invariant  $J_{CP}$, in the quark sector. We use the notation $J_{CP}$ in the neutrino case as well and our prediction for 
 the large range of the $J_{CP}$ is $-0.035~\textrm{to}  ~0.003$,  indicates the amount of the $CP$ violation within the HSMR. 

These observables could be probed further at the ongoing and future experiments like INO, T2K, NO$\nu$A, LBNE, Hyper-K, PINGU and KATRIN \cite{Abe:2011ks,Patterson:2012zs,Adams:2013qkq,Ge:2013ffa,Kearns:2013lea,Athar:2006yb,Drexlin:2013lha}.

  \subsection{Majorana Phases and Their Consequences}\label{majphase}
  We have worked in the CP violating limit which means Majorana and Dirac phases are assumed to be non-zero. Hence,
these non-zero phases are expected to have non-trivial effect on the parameter space. 
One of the main achievements of this work is our predictions regarding the Majorana phases.
One of the observables related to these phases is the $m_{ee}$ and knowing the exact range of the Majorana phases will also allow one to put  upper bound on this parameter~\cite{Delepine}. Determination of the Majorana phases from data on the effective Majorana mass was addressed in~\cite{Pascoli}.
 The correlation  between these phases with $m_{ee}$ is shown in Fig.~\ref{phimee}~\footnote{Posterior plots in this article were made using \texttt{GetDist}~\cite{Lewis:2019xzd}.} and their ranges are given in Table~\ref{tab:fitvalues} (\ref{tab:fitvalues1}) 
at low (high) scale. 
 \begin{figure}[ht]
    \includegraphics[width=.32\textwidth]{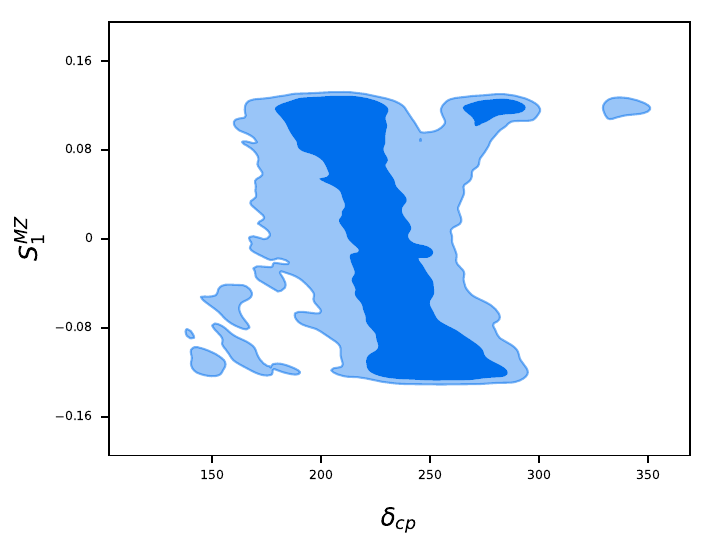}  
  \includegraphics[width=.32\textwidth]{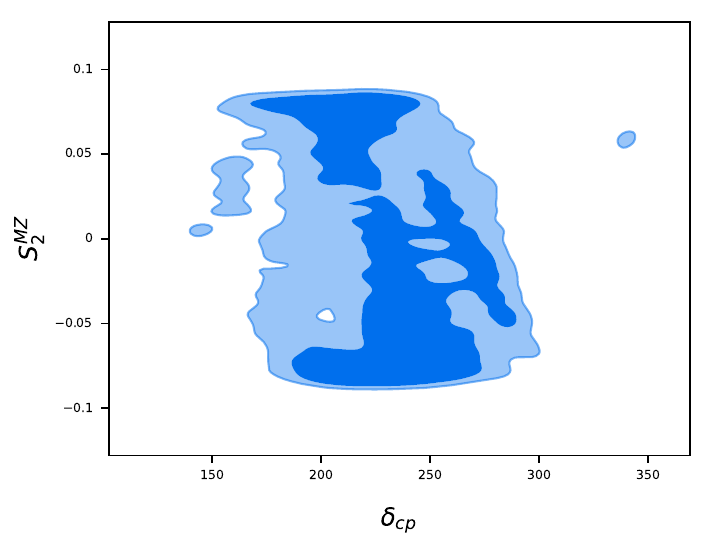}  
 \includegraphics[width=.32\textwidth]{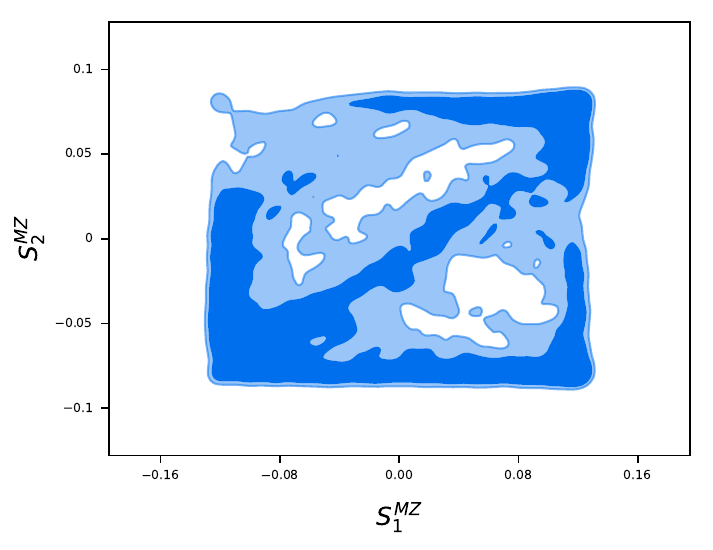}  
 \caption{Posterior distributions of the Majorana neutrino observables $S_1$ and $S_2$ .} \label{s1s2}
\end{figure}

 Although the Dirac phases can bring about the CP-violation irrespective of the nature of the neutrinos,  the
Majorana phases can lead to CP-violation  only if the neutrinos are Majorana particles.
In this part,  we are looking for a signal of CP-violation due to Majorana phases. We know that, Majorana phases can lead to CP-odd effects~\cite{Gouvea} and in fact, 
the present baryon asymmetry in the Universe can be explained with the CP-violation in the decays of a very heavy Majorana neutral leptons, in the  early-universe~\cite{Fukugita}.
 This effect  is CP-odd and comes from the difference between two CP-mirror-image decays.
 An example is the process of “neutrino $\leftrightarrow$ anti-neutrino oscillations”, which was first considered in~\cite{Schechter,Wilczek,Bernabeu,Kayser,Barenboim,Bahcall,Langacker}. It means that the rate for a process of
a neutrino “beam” created by positively-charged leptons in the source and leading to the production
of negatively-charged leptons in the detector

\begin{equation}
l^+_{\alpha}W^-\rightarrow \nu \rightarrow l_{\beta}^-W^+,
\label{osc}
\end{equation}
is different from its CP-mirror image:
\begin{equation}
l^-_{\alpha}W^+\rightarrow \nu \rightarrow l_{\beta}^+W^-,
\label{osc}
\end{equation}

Our analysis allows us to have both the high scale and the low scale ranges of the Majorana phases in our disposal and calculate the CP-odd effect at both scales.
Here we choose the low scale Majorana phases to induce such CP-odd effects. 
Restricting oneself  to the two generation, let say  $2$ and $3$  case,
one finds the CP-odd rate difference to be~\cite{Gouvea}
\begin{equation}
\Delta_{CP}=
\frac{\sin^22\theta_{23}}{E^2}|S|^2m_3m_2
\sin\left(\frac{\Delta m^2_{32}L}{2E}\right)\sin\phi.
\label{DeltaCP}
\end{equation}
Here, $\phi\equiv\phi_2-\phi_1$, S is an additional kinematical factor depending on the initial and final states, which we assume to be of order unity, and $E$ is the energy of the intermediate-state neutrino mass eigenstate which propagates a macroscopic distance $L$. Although, smallness of the ratio $m_2 m_3/ E$ renders $\Delta_{CP}$ unobservable by current and next round experimental set ups like the MINOS long baseline neutrino experiment~\cite{MINOS} etc. .

Among  other observables relating to Majorana phases~\cite{Jarlskog1,Saavedra} we check the invariants 
\begin{eqnarray}
 S_1 &=& Im\left( U_{\nu_e\nu_1}U_{\nu_e\nu_3}^*\right) = \frac{1}{2}
\sin 2\theta_{13}\cos\theta_{12}\sin (\phi_1+\delta) , \nonumber\\
S_2 &=& Im\left( U_{\nu_e\nu_2}U_{\nu_e\nu_3}^*\right) = \frac{1}{2}
\sin 2\theta_{13}\sin\theta_{12}\sin (\phi_2+\delta)\, .  \nonumber
\end{eqnarray}

The correlations between these variables and $\delta_{CP}$, plotted in Fig.~\ref{s1s2}, are presented for the first time here.  
These are very interesting because, 
knowing the value of one of them, let say $S_1$, will directly allow us to know the value of $S_2$. Our prediction for these parameters at low scale are  $-0.125<S_1<0.126$ and $-0.0856 <S_2< 0.00849$.

It is worthwhile to mention that,  the confirmation of the strong correlations among different observables, which we have provided so far,  would be a strong signal of an underlying dynamics
 having a common origin of quark-lepton mixing. It could be for  example the quark-lepton unification or flavour symmetries or both. 

\section{Summary and Outlook} 
\label{conclusions}
For an insight towards finding the origin of the remarkable difference between the mixing parameters in neutrino and quark sectors, the high-scale mixings relations or unification hypotheses were developed and
the free parameters were analysed with respect to experiments. In this article we have introduced
a Bayesian exploration technique for extending or complementing previous explorations of model parameters
based on the mentioned hypotheses.

Finally, in short, crux of our paper is following.

\begin{itemize}

\item
Our analysis suggest that HSMR  provides a very simple explanation of the observed large neutrino mixing. This is a corollary of previous studies 
of Majorana neutrinos without the Majorana phases~\cite{Abbas:2015vba},  and Dirac neutrinos~\cite{Abbas-Abyaneh2} . In this work, taking into account the Majorana phases we conclude that,
 the HSMU hypothesis is not compatible with the existing bounds on the mixing parameters.

\item
We have done a rigorous, thorough and comprehensive study of the HSMR via the Bayesian exploration technique using the REAP~\cite{reap} 
and MultiNest~\cite{Skilling,Feroz,multinest} codes, which is done for the first time.
This technique can readily be
used for further comparisons and analyses of various high-scale mixing relations, such as in \cite{Abbas:2015vba}, with respect to experimental data or constraints.  

\item
We have also discovered new strong correlations among different experimental observables. Apart from correlations for $\theta_{12}$,  $\theta_{13}$ and  $\theta_{23}$,  which
 are easily testable in present ongoing experiments, our prediction for $\delta_{cp}$ is a subset of the global fit result and the 
range for $J_{CP}$ to be  $-0.035~\textrm{ to } ~0.003$, indicates the amount of the $CP$ violation in our analysis, which can be tested.

\item
Our bound for $M_{ee}$ at $95\%$ confidence region is $0.019 < M_{ee}<0.056$ with a central value at $0.042$. This is a very precise prediction compared to both GERDA, $M_{ee}< 0.182$~\cite{Gerda2}  and Kamland~\cite{Kamland} $ M_{ee} < 0.168$ that can be tested in the next runs of these experiments.  Contrary to previous studies on this subject, our bounds on $M_{e}$ and $\Sigma m_i$, are totally in agreement with 
the experimental bounds from KATRIN~\cite{Aseev:2011dq}-~\cite{KATRIN0} and cosmological observations~\cite{Ade:2015xua}, respectively and are very precise.

\item
One of the main achievements of this work is to find strong correlations between the  Majorana phases or their functions with respect to other quantities via HSMR, many of which appear in this paper for the first time.
In this respect, the posterior distribution of  Majorana phases with respect to $m_{ee}$ is shown in Fig.~\ref{phimee}. 
\item
We have calculated the $\Delta_{CP}$  as a signal of CP-violation due to Majorana phases by computing the rate for the process of “neutrino $\leftrightarrow$ anti-neutrino oscillation”. Any measurement for $\Delta_{CP}$ can be a direct test for our prediction of the value of the Majorana phases.

\item
The correlations between the $S_1$ and $S_2$ invariants as functions of the Majorana phases, with respect to $\delta_{CP}$ are plotted in Fig.~\ref{s1s2} and are presented for the first time here.  These are very interesting because, measuring one of them,  we only need to measure $\delta_{CP}$, $ \theta_{13}$ and $\sin\theta_{12}$ to know the value of the Majorana phases and knowing the value of one of them, let say $S_1$, will directly allow us to know the value of $S_2$.
\item
We have done the analysis for the case when no threshold corrections are included  and results are presented.
\item
Our results for numerical values of the leptonic mixing angles at the high scales, specifically the $\alpha_i$s, which are given in table~\ref{tab:fitvalues1}, could be used as a guide for leptonic mixings model building and also to constrain the theory space of HSMR. A more detailed and elaborated answer to this comment requires a through investigation, that is beyond the scope of this paper and will be dealt with in a separate work.
\item
The packages developed for this work is publicly available at \hyperlink{https://github.com/shehu0/NeutrinoPh}{https://github.com/shehu0/NeutrinoPh}.

\end{itemize}

\section*{Acknowledgements}
\indent
M. Z. A wants to thank Gauhar Abbas his very helpful discussions. 


\begin{thebibliography}{99}

\bibitem{Pati:1974yy}
  J.~C.~Pati and A.~Salam,
  Phys.\ Rev.\ D {\bf 10} (1974) 275
   [Erratum-ibid.\ D {\bf 11} (1975) 703].


\bibitem{Georgi:1974sy}
  H.~Georgi and S.~L.~Glashow,
  Phys.\ Rev.\ Lett.\  {\bf 32} (1974) 438.


\bibitem{Fritzsch:1974nn}
  H.~Fritzsch and P.~Minkowski,
  Annals Phys.\  {\bf 93} (1975) 193.



\bibitem{Altarelli:2010gt}
  G.~Altarelli and F.~Feruglio,
  Rev.\ Mod.\ Phys.\  {\bf 82} (2010) 2701
  [arXiv:1002.0211 [hep-ph]].

\bibitem{King:2013eh}
  S.~F.~King and C.~Luhn,
  Rept.\ Prog.\ Phys.\  {\bf 76} (2013) 056201
  [arXiv:1301.1340 [hep-ph]].


\bibitem{Holthausen:2013vba}
  M.~Holthausen and K.~S.~Lim,
  Phys.\ Rev.\ D {\bf 88} (2013) 033018
  [arXiv:1306.4356 [hep-ph]].

\bibitem{Araki:2013rkf}
  T.~Araki, H.~Ishida, H.~Ishimori, T.~Kobayashi and A.~Ogasahara,
  Phys.\ Rev.\ D {\bf 88} (2013) 096002
  [arXiv:1309.4217 [hep-ph]].

\bibitem{Ishimori:2014jwa}
  H.~Ishimori and S.~F.~King,
  Phys.\ Lett.\ B {\bf 735} (2014) 33
  [arXiv:1403.4395 [hep-ph]].

\bibitem{Lam:2014kga}
  C.~S.~Lam,
  Phys.\ Rev.\ D {\bf 89} (2014) 9,  095017
  [arXiv:1403.7835 [hep-ph]].







  A.~Y.~Smirnov,
  hep-ph/0402264.

\bibitem{Raidal:2004iw}
  M.~Raidal,
  Phys.\ Rev.\ Lett.\  {\bf 93} (2004) 161801
  [hep-ph/0404046].

\bibitem{Minakata:2004xt}
  H.~Minakata and A.~Y.~Smirnov,
  Phys.\ Rev.\ D {\bf 70} (2004) 073009
  [hep-ph/0405088].

\bibitem{Frampton:2004vw}
  P.~H.~Frampton and R.~N.~Mohapatra,
  JHEP {\bf 0501} (2005) 025
  [hep-ph/0407139].

\bibitem{Kang:2005as}
  S.~K.~Kang, C.~S.~Kim and J.~Lee,
  Phys.\ Lett.\ B {\bf 619} (2005) 129
  [hep-ph/0501029].



\bibitem{Abe:2011sj}
  K.~Abe {\it et al.}  [T2K Collaboration],
  Phys.\ Rev.\ Lett.\  {\bf 107} (2011) 041801
  [arXiv:1106.2822 [hep-ex]].

\bibitem{Adamson:2011qu}
  P.~Adamson {\it et al.}  [MINOS Collaboration],
  Phys.\ Rev.\ Lett.\  {\bf 107} (2011) 181802
  [arXiv:1108.0015 [hep-ex]].


\bibitem{Abe:2012tg}
  Y.~Abe {\it et al.}  [Double Chooz Collaboration],
  Phys.\ Rev.\ D {\bf 86} (2012) 052008
  [arXiv:1207.6632 [hep-ex]].

\bibitem{Ahn:2012nd}
  J.~K.~Ahn {\it et al.}  [RENO Collaboration],
  Phys.\ Rev.\ Lett.\  {\bf 108} (2012) 191802
  [arXiv:1204.0626 [hep-ex]].

\bibitem{An:2012eh}
  F.~P.~An {\it et al.}  [Daya Bay Collaboration],
  Phys.\ Rev.\ Lett.\  {\bf 108} (2012) 171803
  [arXiv:1203.1669 [hep-ex]].


\bibitem{Antusch:2012fb}
  S.~Antusch, C.~Gross, V.~Maurer and C.~Sluka,
  Nucl.\ Phys.\ B {\bf 866} (2013) 255
  [arXiv:1205.1051 [hep-ph]].






\bibitem{Abbas:2014ala}
  G.~Abbas, S.~Gupta, G.~Rajasekaran and R.~Srivastava,
  Phys.\ Rev.\ D {\bf 89} (2014) 9,  093009
  [arXiv:1401.3399 [hep-ph]].


\bibitem{Abbas:2013uqh} 
  G.~Abbas, S.~Gupta, G.~Rajasekaran and R.~Srivastava,
  Phys.\ Rev.\ D {\bf 91}, no. 11, 111301 (2015)
  doi:10.1103/PhysRevD.91.111301
  [arXiv:1312.7384 [hep-ph]].


\bibitem{Srivastava:2015tza} 
  R.~Srivastava,
  Pramana {\bf 86}, no. 2, 425 (2016)
  [arXiv:1503.07964 [hep-ph]].


\bibitem{Srivastava:2016fhg} 
  R.~Srivastava,
  Springer Proc.\ Phys.\  {\bf 174}, 369 (2016).



\bibitem{Haba:2012ar} 
  N.~Haba and R.~Takahashi,
  Europhys.\ Lett.\  {\bf 100}, 31001 (2012)
  doi:10.1209/0295-5075/100/31001
  [arXiv:1206.2793 [hep-ph]].





\bibitem{Mohapatra:2005gs}
  R.~N.~Mohapatra, M.~K.~Parida and G.~Rajasekaran,
  Phys.\ Rev.\ D {\bf 71} (2005) 057301
  [hep-ph/0501275].

\bibitem{Mohapatra:2005pw}
  R.~N.~Mohapatra, M.~K.~Parida and G.~Rajasekaran,
  Phys.\ Rev.\ D {\bf 72} (2005) 013002
  [hep-ph/0504236].


\bibitem{Agarwalla:2006dj}
  S.~K.~Agarwalla, M.~K.~Parida, R.~N.~Mohapatra and G.~Rajasekaran,
  Phys.\ Rev.\ D {\bf 75} (2007) 033007
  [hep-ph/0611225].

\bibitem{Abbas:2015vba} 
  G.~Abbas, M.~Z.~Abyaneh, A.~Biswas, S.~Gupta, M.~Patra, G.~Rajasekaran and R.~Srivastava,
  Int.\ J.\ Mod.\ Phys.\ A {\bf 31}, no. 17, 1650095 (2016)
  doi:10.1142/S0217751X16500950
  [arXiv:1506.02603 [hep-ph]].
  
\bibitem{Abbas-Abyaneh2} 
G. Abbas, M. Z. Abyaneh, R. Srivastava, Phys.\ Rev.\ D {\bf 95}, 075005 (2017)

\bibitem{BilenkyHosek} S. M. Bilenky, J. Hosek and S. T. Petcov, “On Oscillations of Neutrinos with Dirac and Majorana Masses,” Phys. Lett. 94B (1980) 495. doi:10.1016/0370-2693(80)90927-2
\bibitem{GirardiPetcov}I. Girardi, S. T. Petcov and A. V. Titov, “Predictions for the Majorana CP Violation Phases
in the Neutrino Mixing Matrix and Neutrinoless Double Beta Decay,” Nucl. Phys. B 911
(2016) 754 doi:10.1016/j.nuclphysb.2016.08.019 [arXiv:1605.04172 [hep-ph]].
\bibitem{Weinbergscale} J. H.Garcia, M. A. Schmidt,  Eu.\ Phys.\ J.\ C  {\bf79}, 938 (2019)

\bibitem{Alessandria:2011rc} 
  F.~Alessandria, E.~Andreotti, R.~Ardito, C.~Arnaboldi, F.~T.~Avignone, III, M.~Balata, I.~Bandac and T.~I.~Banks {\it et al.},
  arXiv:1109.0494.

\bibitem{Auger:2012ar} 
  M.~Auger {\it et al.}  [EXO Collaboration],
  Phys.\ Rev.\ Lett.\  {\bf 109}, 032505 (2012),
  arXiv:1205.5608.


\bibitem{Minkowski} 
  P.~Minkowski,
  Phys.\ Lett.\ B {\bf 67}, 421 (1977).

\bibitem{GellMann:1980vs} 
  M.~Gell-Mann, P.~Ramond and R.~Slansky,
  Conf.\ Proc.\ C {\bf 790927}, 315 (1979),
  arXiv:1306.4669.

\bibitem{Yanagida:1979as} 
  T.~Yanagida,
  Conf.\ Proc.\ C {\bf 7902131}, 95 (1979).

\bibitem{Glashow:1979nm} 
  S.~L.~Glashow,
  NATO Adv.\ Study Inst.\ Ser.\ B Phys.\  {\bf 59}, 687 (1980).

\bibitem{Mohapatra:1979ia} 
  R.~N.~Mohapatra and G.~Senjanovic,
  Phys.\ Rev.\ Lett.\  {\bf 44}, 912 (1980).




\bibitem{Casas:2003kh} 
  J.~A.~Casas, J.~R.~Espinosa and I.~Navarro,
  JHEP {\bf 0309}, 048 (2003)
  [hep-ph/0306243].

\bibitem{Antusch:2005gp0} S. Antusch, M. Drees, J. Kersten, M. Lindner and M. Ratz, Phys.\ Lett. \ B {\bf 519}, 238
(2001), arXiv:hep-ph/0108005.

\bibitem{Drees2} S. Antusch, M. Drees, J. Kersten, M. Lindner and M. Ratz, Phys. Lett. B 525, 130
(2002), arXiv:hep-ph/0110366.
\bibitem{Antusch:2005gp} 
  S.~Antusch, J.~Kersten, M.~Lindner, M.~Ratz and M.~A.~Schmidt,
  JHEP {\bf 0503}, 024 (2005), hep-ph/0501272.
  \bibitem{Pluciennik} P. H. Chankowski and Z. Pluciennik, Phys.\ Lett.\ B {\bf 316}, 312 (1993), arXiv:hep-
ph/9306333.
\bibitem{Babu} K. S. Babu,  C. N. Leung and J. T. Pantaleone, Phys.\ Lett.\ B {\bf 319}, 191 (1993),
arXiv:hep-ph/9309223.
 \bibitem{Antusch0} S. Antusch, J. Kersten, M. Lindner and M. Ratz, Nucl.\ Phys.\ B {\bf 674}, 401 (2003),
arXiv:hep-ph/0305273.

\bibitem{Casas:1999tg} 
  J.~A.~Casas, J.~R.~Espinosa, A.~Ibarra and I.~Navarro,
  Nucl.\ Phys.\ B {\bf 573}, 652 (2000)
  doi:10.1016/S0550-3213(99)00781-6
  [hep-ph/9910420].
\bibitem{Mohapatramodel}R. N. Mohapatra, M. K. Parida and G. Rajasekaran, Phys. \ Rev. \ D {\bf 69} 053007 (2004)  [hep-
ph/0301234].

\bibitem{Chun1} E. J. Chun and S. Pokorski, Phys.\ Rev. \ D {\bf 62}, 053001 (2000) [hep-ph/9912210].
\bibitem{Chankowski1}  P. H. Chankowski, A. Ioannisian, S. Pokorski and J. W. FValle, Phys.\  Rev. \ Lett. \ {\bf 86}, 3488
(2001) [hep-ph/0011150].
\bibitem{Chun2}  E. J. Chun, Phys.\ Lett.\ B {\bf 505}, 155 (2001) [hep-ph/0101170].
\bibitem{Chankowski2}  P. H. Chankowski and P. Wasowicz, Eur. \ Phys.\ J. \ C {\bf 23}, 249 (2002) [hep-ph/0110237].




\bibitem{Gerda1} 
  M.~Agostini {\it et al.}  [GERDA Collaboration],
  Phys.\ Rev.\ Lett.\  {\bf 120}, 132503 (2018),
 arXiv:1803.11100.
\bibitem{Gerda2} XXIX Int. "Conference on Neutrino Physics and Astrophysics", June 22 - July 1, (2020).

\bibitem{Kamland} 
  A.~Gando {\it et al.}  [KamLAND-Zen Collaboration],
  Phys.\ Rev.\ Lett.\  {\bf 117}, 082503 (2016).


\bibitem{Globalfit} I. Esteban, M.C. Gonzalez-Garcia, A. Hernandez-Cabezudo, M. Maltoni, T. Schwetz, 	JHEP {\bf 01}, 106 (2019), arXiv:1811.05487.

\bibitem{reap} 
 S. Antusch, J. Kersten, M. Lindner, M. Ratz and M. A. Schmidt, J.\ High Energy
Phys.\ {\bf 0503}, 024 (2005), arXiv:hep-ph/0501272.


  \bibitem{Skilling}
  J.~Skilling,
  AIP Conference Proceedings of the 24th International 
  Workshop on Bayesian Inference and Maximum Entropy Methods in Science and 
  Engineering, Vol. 735, pp. 395-405 (2004)
  available from http://www.inference.phy.cam.ac.uk/bayesys/


\bibitem{Feroz}
  F.~Feroz and M.~P.~Hobson
  {\em  Multimodal nested sampling: an efficient and robust alternative to 
  MCMC methods for astronomical data analysis},
  Mon.\ Not.\ Roy.\ Astron.\ Soc. {\bf 384} (2008) 449
  [arXiv:0704.3704].



\bibitem{multinest} F. Feroz, M.P. Hobson, M. Bridges, Mon.\ Not.\ Roy.\ Astron.\ Soc.\ {\bf 398} 1601(2009), 	arXiv:0809.3437.
  

  
 
 \bibitem{Rodejohann} W. Rodejohann, J.\ Phys.\ G {\bf 39}, 124008 (2012), arXiv:1206.2560 [hep-ph].
 \bibitem{Bilenky}  S. M. Bilenky and C. Giunti, Int.\ J.\ Mod.\ Phys.\ A {\bf 30}, 1530001 (2015), arXiv:1411.4791.
   \bibitem{Bilenky1}S.~M.~Bilenky, C. Giunti, C.W. Kim, S.T. Petcov, Phys.\ Rev.\ D {\bf 54} , 4432 (1996), hep-ph/9604364.
   \bibitem{Bilenky2} S. M. Bilenky, S. Pascoli, S. T. Petcov, Phys.\ Rev.\ D {\bf 64}, 053010 (2001), hep-ph/0102265.

\bibitem{Aseev:2011dq}
  V.~N.~Aseev {\it et al.}  [Troitsk Collaboration],
  Phys.\ Rev.\ D {\bf 84} (2011) 112003
  [arXiv:1108.5034 [hep-ex]].
  
  

\bibitem{Kraus:2004zw}
  C.~Kraus, B.~Bornschein, L.~Bornschein, J.~Bonn, B.~Flatt, A.~Kovalik, B.~Ostrick and E.~W.~Otten {\it et al.},
  Eur.\ Phys.\ J.\ C {\bf 40} (2005) 447
  [hep-ex/0412056].

\bibitem{Mainz} J. Bonn et al., The Mainz neutrino mass experiment, Nucl. Phys. Proc. Suppl. 91 (2001) 273–279. [PoShep2001,192(2001)].  
\bibitem{Drexlin:2013lha} 
  G.~Drexlin, V.~Hannen, S.~Mertens and C.~Weinheimer,
  Adv.\ High Energy Phys.\  {\bf 2013}, 293986 (2013), arXiv:1307.0101.
 
  \bibitem{KATRIN0} M. Aker et. al. , "An improved upper limit on the neutrino mass from a direct kinematic method by KATRIN", 	arXiv:1909.06048.

  
\bibitem{Lewis:2019xzd}
  A.~Lewis,
  arXiv:1910.13970 [astro-ph.IM].


\bibitem{Ade:2015xua}
  P.~A.~R.~Ade {\it et al.}  [Planck Collaboration],
  arXiv:1502.01589 [astro-ph.CO].
\bibitem{Krastev-cp}
P.~I.~Krastev and S.~T.~Petcov,
Phys. Lett. B \textbf{205} (1988), 84-92
doi:10.1016/0370-2693(88)90404-2






  


\bibitem{Athar:2006yb} 
  M.~S.~Athar {\it et al.}  [INO Collaboration],
  INO-2006-01.


\bibitem{Abe:2011ks} 
  K.~Abe {\it et al.}  [T2K Collaboration],
  Nucl.\ Instrum.\ Meth.\ A {\bf 659}, 106 (2011), arXiv:1106.1238.


\bibitem{Patterson:2012zs} 
  R.~B.~Patterson [NOvA Collaboration],
  Nucl.\ Phys.\ Proc.\ Suppl.\  {\bf 235-236}, 151 (2013), arXiv:1209.0716.


\bibitem{Adams:2013qkq} 
  C.~Adams {\it et al.}  [LBNE Collaboration],
  arXiv:1307.7335.


\bibitem{Kearns:2013lea} 
  E.~Kearns {\it et al.}  [Hyper-Kamiokande Working Group Collaboration],
  arXiv:1309.0184.
  
\bibitem{Ge:2013ffa} 
  S.~-F.~Ge and K.~Hagiwara,
  arXiv:1312.0457.
      \bibitem{Delepine} D. Delepine, V. G. Macias, S. Khalil, G. L. Castro, \ Phys.  \ Lett. {\bf B 693}, 438 (2010)
  \bibitem{Pascoli} S. Pascoli et al.,  Nucl.\ Phys.\ B {\bf 734}, 24 (2006) arXiv: 0505226.
  \bibitem{Gouvea}
  A. Gouvea, B. Kayser, R. N. Mohapatra,  Phys.\ Rev.\ B {\bf 67}, 053004 (2003)
  \bibitem{Fukugita} M. Fukugita and T. Yanagida, \ Phys. \ Lett. {\bf B174}, 45 (1986).
  \bibitem{Schechter} J. Schechter and J. F. Valle, \ Phys. \ Rev. {\bf D 23}, 1666 (1981).
  \bibitem{Wilczek} L. F. Li and F. Wilczek, \ Phys. \ Rev. {\bf D 25}, 143 (1982).
  \bibitem{Bernabeu}J. Bernabeu and P. Pascual,\ Nucl. \ Phys.{\bf B 228}, 21 (1983).
  \bibitem{Kayser} B. Kayser, \ Phys.\ Rev.{\bf D 30}, 1023 (1984).
  \bibitem{Barenboim}G. Barenboim, J. Bernabeu and O. Vives, \ Phys.\  Rev. {\bf Lett.  77}, 3299 (1996).
  \bibitem{Bahcall} J. N. Bahcall and H. Primakoff, \ Phys. \ Rev. {\bf D 18}, 3463 (1978).
  \bibitem{Langacker} P. Langacker and J. Wang, \ Phys.\ Rev. {\bf D 58}, 093004 (1998).
    \bibitem{MINOS} MINOS Collaboration, \ Phys. \ Rev. \ Lett. {\bf110} , 251801 (2013)
   \bibitem{Jarlskog1} C. Jarlskog, \ Phys. \ Rev. \ Lett. {\bf 55}, 1039 (1985).
   \bibitem{Saavedra} J. A. Aguilar-Saavedra and G. C. Branco, \ Phys. \ Rev. {\bf D 62}, 096009 (2000)



 
 

\end{thebibliography}
\end{document}